# A Taxonomy of Data Quality Challenges in Empirical Software Engineering

Michael Franklin Bosu and Stephen G. MacDonell

*SERL, School of Computing and Mathematical Sciences*
*Auckland University of Technology*
*Auckland, New Zealand*
michael.bosu@aut.ac.nz, stephen.macdonell@aut.ac.nz

**Abstract**

*Reliable empirical models such as those used in software effort estimation or defect prediction are inherently dependent on the data from which they are built. As demands for process and product improvement continue to grow, the quality of the data used in measurement and prediction systems warrants increasingly close scrutiny. In this paper we propose a taxonomy of data quality challenges in empirical software engineering, based on an extensive review of prior research. We consider current assessment techniques for each quality issue and proposed mechanisms to address these issues, where available. Our taxonomy classifies data quality issues into three broad areas: first, characteristics of data that mean they are not fit for modeling; second, data set characteristics that lead to concerns about the suitability of applying a given model to another data set; and third, factors that prevent or limit data accessibility and trust. We identify this latter area as of particular need in terms of further research.*

## 1. BACKGROUND AND MOTIVATION

Measurement data are used to support many aspects of software development and management, but effort estimation and defect classification are particularly prevalent uses of such data. The empirical software engineering (ESE) community is unfortunately not immune to studies that have used questionable or poor quality measurement data in model-building [1–4]. Quality is defined here as being "fit for purpose" [5] and we expect, we *need* ESE data to be of high quality – to be fit for the modeling task at hand, be it for classification or prediction.

A study conducted by Bobrowski et al. [6] presented what they described as a software engineering view of data quality, comprising three perspectives: data quality metrics, data testing and data quality requirements. Their work considered quality in terms of the data used or accessed by software systems. In contrast, in this paper we propose a taxonomy that addresses data quality from the point of view of the data used in modeling phenomena concerning the software process or product. Of course, data quality problems are not limited to the software engineering domain, and our taxonomy has been informed by relevant work in other disciplines. Du & Zhou [7], for instance, created a taxonomy of data quality problems for online financial data and proposed an ontology-based framework to improve the quality of online-financial data. Three elements of their framework (unreliable data, inconsistent representation and missing data) [7] can be mapped to three elements of the accuracy class of our taxonomy (described in detail in section III).

The taxonomy is intended to be useful to both researchers and practitioners by bringing to the attention of the entire ESE community the potential problems that can exist in the data sets we work with, or that are derived from our work, along with some current treatments or solutions. As the taxonomy captures the major data quality challenges that affect ESE data sets, empirical software engineering researchers and practitioners alike can benchmark their data quality assessment against it, to identify any potential sources of error or unsuitability for modeling and management. Considering the fact that ESE researchers work predominantly with secondary data it is essential to be aware of the nature of the data that exists, especially in the public domain. The quality of ESE data sets cannot be taken for granted, as data collected even by highly mature organizations can have issues. This is evident in the discovery by Gray et al. [8] of several data quality problems with the NASA Metrics Program data sets that are used widely for defect prediction research. The issues evident in these data sets are several, and include redundant data, inconsistencies, constant attribute values, missing values and noise.

In a more general commentary Liebchen & Shepperd [9] bemoan the lack of interest by the software engineering community in addressing the issue of data quality. They looked in depth at the data quality dimension of accuracy, with accuracy being defined as the absence of noise. They were surprised by the few studies in ESE that had considered data quality explicitly [9]. Due in part to this lack of interest there have been many instances where models have been built without any form of preprocessing or quality checks on the data. Thus, while much attention has been given to the development of prediction systems, the same cannot be said regarding the quality of the data used in generating those systems. We therefore set out to identify all of the data quality issues associated with the collection and use of ESE data sets. Our goal is to first improve awareness and understanding of the diverse data quality issues that can arise in ESE, so second we may improve both the quality of ESE research and the practice of software engineering.

The taxonomy presented here therefore captures the many challenges associated with data typically used in ESE modeling. Although some of the elements of the taxonomy might not be peculiar to ESE data sets, to the best of our knowledge they have not been addressed sufficiently in other domains to enable ESE researchers to readily borrow solutions developed in those domains.

The rest of the paper is organized as follows. In section II we present background information on ESE data. In section III we present the taxonomy based on a comprehensive review of prior research in ESE, but also informed by considerations of data quality in other disciplines. In section IV we consider the implications for research in a discussion of the taxonomy, and we then conclude the paper in section V with recommendations for research and practice.

## 2. EMPIRICAL SOFTWARE ENGINEERING DATA

'Software metrics' is the collective term commonly used to describe the wide range of activities concerned with measurement in software engineering. These activities range from producing numbers that characterize properties of software code (these are the classic software metrics) through to models that help predict software resource requirements and software quality. The subject area also includes the quantitative aspects of quality control and assurance - and this covers activities such as recording and monitoring defects during development and testing [10]. The use of software metrics is generally accepted as a means of supporting rational decision making during software development and maintenance [11][12], with broader goals of increased productivity and quality and reduced cycle time [13]. Metrics have been designed and are used to measure a diverse set of product, process and resource characteristics, including system size, software quality, development schedule, developer effort and code complexity [12]. The quality of software engineering data has been questioned due to known problems with the collection of the data (as detailed below), but its trustworthiness is also at issue as there are perceptions that data are usually massaged by managers so that it appears better than the true reality [10].

The quality of data used in empirical software engineering can be improved at multiple points in the process, including at the collection stage. Johnson & Disney [1] questioned the quality of data collection related to the personal software process (PSP). PSP data in their study was recorded manually by students and then verified – also manually – by instructors for accuracy. Upon examining the data the instructors identified errors that represented impossible combinations of data. Their results raised questions about the accuracy of (their) manually collected and analyzed PSP data. They proposed the use of integrated tool support for higher quality PSP data. The challenge inherent in the PSP is that the developer collects data about his/her own work practices, such as effort expended on tasks. This has a distinct advantage over other approaches such as collection or estimation by others (perhaps managers), or automated collection, in placing the responsibility for correct collection with the developer. On the downside, this can lead to work overload for the developer, and there is also the question of developer honesty. Errors of omission, addition and transcription were identified by Johnson & Disney as occurring at the data collection stage. Such errors were noted as being the most difficult to reproduce and resolve because time has generally passed before they are detected (if indeed they are detected at all).

Software engineering (SE) is a technical domain, but those who submit or provide data to improve the practice of SE might not necessarily be SE professionals such as programmers, and so could submit data without understanding its implications. These submitters, lacking domain knowledge, might not even be in a position to check the validity of the data they submit. For instance, bug reports are often submitted by users who are unlikely to be in a position to assess the quality or veracity of their reports, and/or they may not be aware of other reports that have reported the same or similar issues. Yet we use metrics such as numbers of bug reports, sometimes without question, in our quality modeling.

In their study gauging the acceptance of software metrics in a large multinational organisation, Umarji & Seaman [14] identified various challenges in relation to the use of metrics from the perspective of developers and managers, noting in particular the inflexibility of the metric tool and the lack of expertise within the team. In this instance the tool had very limited functionality as well as poor support for fixing bugs and adding new features - it was basically a time reporting tool and was not flexible enough to be tailored to different types of metrics (to the extent that new measures had to be collected in a separate spreadsheet). According to the managers involved, lack of expertise affected the appropriate definition and identification of risks and increased the resistance to the adoption of common terminology, meaning that the data collected was not understood in a consistent way.

This can be contrasted with the approach reported by Staron et al. [15] in relation to the introduction of a framework for developing measurement systems at Ericsson. A dedicated person/role was assigned responsibility to present data and to prevent it from being biased; that person was able to explain in detail how the data were collected, processed and presented to ensure that the numbers reflected reality.

In order to make the auditing of the data collection stage more transparent and so increase the trustworthiness of the data, we propose the use of a provenance system to record and potentially replay collection procedures when discrepancies are identified. Provenance is addressed in detail in the next section.

## 3. DATA QUALITY TAXONOMY

We surveyed a decade of recent literature on data quality in ESE and identified a range of quality issues. We then grouped these issues into three main classes in the proposed taxonomy. First is the group of characteristics of data that mean the observations are not fit for model-building (accuracy); second are data set characteristics that lead to concerns about the suitability of applying one model to another data set (relevance); and third is a set of factors that limit data accessibility and trust (provenance). Our intention is to use this taxonomy to bring to the attention of the wider ESE community the challenges associated with data used in modeling and the techniques that have been proposed to identify and/or resolve some of these problems. We do not provide detailed coverage of all relevant studies here due to the fact that some of the studies grouped under the individual elements of the taxonomy have similar themes. Rather, we present in this section a representative set of studies that serve to illustrate the elements of the taxonomy. (All studies reviewed have been considered in the Discussion, however (section IV).) Figure 1 depicts the proposed taxonomy.

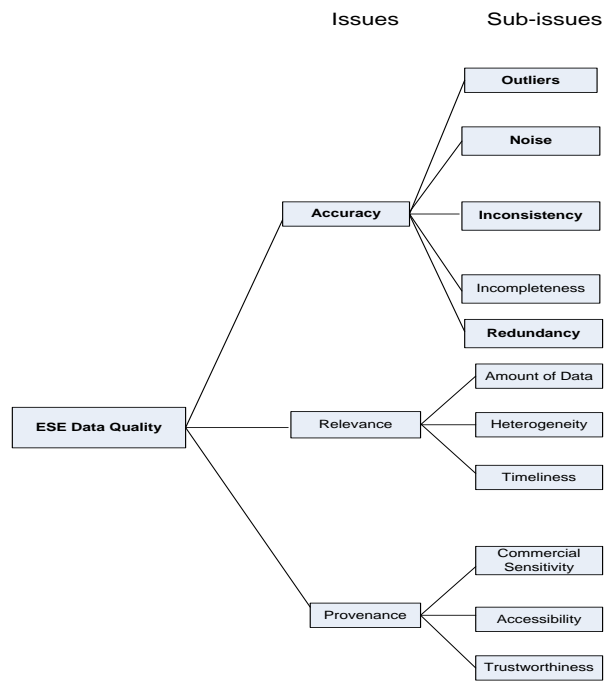

Figure 1. Taxonomy of Data Quality in ESE.

### A. Accuracy

The Oxford English Dictionary defines accuracy as the state of being accurate; precision or exactness resulting from care; hence precision …exactness, correctness. In ESE, accuracy as it relates to data means the correctness of the data or the absence of noise. Data accuracy is essential to ESE research in particular but also to the broader discipline of software engineering practice, since researchers rely on data to design and create classification and prediction models to improve the practice of software engineering. If there are underlying quality problems with the inputs to a model, then the resultant model cannot be expected to provide outcomes that software practitioners will use.

*1) Noise*

Noise – erroneous data – has been identified in several software measurement data sets [1], [2], [16] and the ESE community has responded with a number of studies that seek to address the incidence and effects of noise.

Liebchen et al. [3] conducted classification experiments to assess the effect that noise has on predictive accuracy and to evaluate the robustness and accuracy of techniques for handling noise in ESE data sets. Three noise correction techniques were employed: robust algorithms, filtering, and polishing. Their results demonstrated that polishing improves classification accuracy more effectively compared to noise elimination and robust algorithm approaches.

Khoshgoftaar et al. applied several noise detection and correction procedures to ESE data sets across a range of studies [4], [16–18], with varying degrees of success. Noise detection techniques including Bayesian multiple imputation, a clustering-based noise detection approach using the k-means algorithm, an Ensemble-Partition filter, a technique to detect noise relative to an attribute of interest (AOI), rule-based noise detection and Closest List Noise Identification were applied to various ESE data sets. However, variations in noise characteristics mean that it is difficult to settle on a technique that can be said to be a best fit for all noise types. Furthermore, while the removal of noisy instances or the application of robust algorithms has been shown to improve model performance [19–21], these various studies also highlight the relative immaturity of the community's work on noise detection and correction, in that all manner of filters and algorithms are being tried with little clear dominance of any particular approach.

*2) Outliers*

Outliers have been noted as a frequent phenomenon in software measurement data [1], [3], [22], a consequence of the skewed distributions that often result from metric data collection efforts. An outlier is an observation that lies outside the overall pattern of a distribution [23]. Usually, the presence of an outlier indicates some sort of problem: this can be a case that does not fit the model under study or an error in measurement.

In this paper, we adopt the definition of outlier attributed to Yoon & Bae who define an outlier as a project instance with one or more abnormal attributes [24]. Yoon & Bae proposed a pattern-based outlier detection method that identifies abnormal attributes in software project data: after discovering reliable and frequent patterns that reflect the typical characteristics of software project data, outliers and their abnormal attributes can be detected by matching the software project data with those patterns [24]. The Yoon & Bae study is significant in the sense that the abnormality of outliers is determined and acted upon relative to other data, in contrast to many studies that classify all outliers as noise and simply remove them. Although pattern-based outlier detection holds some promise, we do not have underlying theories that tell us what patterns to expect, and at times we may be vulnerable to the use of heterogeneous data sets that mask patterns relevant to particular subsets.

Statistical analysis is also commonly used to identify outliers [25], [26]. Three mechanisms have been frequently employed to deal with outliers (or not) in ESE. First is a do-nothing approach that leaves the outlier instance in the data set - this could be due to the fact that removal of the instance might mean the model would not be statistically significant. This in itself is not a sound reason for ongoing inclusion and this approach is not recommended. Second, and the predominant practice, is the removal of outliers with the justification being that they are extreme observations. Again this is not ideal as outliers may well be valid, albeit unusual, observations. Third, robust algorithms such as least-median squares regression [27] and Bayesian nets, which are resistant to outliers, have been employed to mitigate the above two situations [28]. For instance, Lavazza & Morasca [29] used a robust regression method – a generalization of the Least Median of Squares – in order not to discard too many data points due to outliers, because as much as 57% of the data points in one of their data sets were determined to be outliers from a Least Squares perspective. Outliers have been a constant source of problems in the analysis of ESE data [27]. In some cases, outliers are due to corrupted data, while they may be the result of highly unlikely circumstances in others. Outliers may significantly bias regression coefficients when using standard least squares regression methods. In order to prevent this condition, Abrahamsson et al. [30] also employed robust regression techniques when they performed iterative effort prediction. Another approach that could be used to address outliers is to allow the outlier instance(s) to remain in the data set but to model the data as two (or more) distinct distributions.

Such an approach was appropriate in the study of commits in source control repositories reported by Hindle et al. [31]. Although large commits are often considered as outliers, the authors demonstrated that in many cases they were fundamentally important to the resulting software architecture. In short, outliers may or may not be problematic in any given case, and therefore analysis is needed to confirm their existence, the cause of their existence, and their potential effect on any models generated. Blanket discarding of unusual data points is ill-advised.

*3) Incompleteness*

It is widely acknowledged that software engineering measurement data sets are often affected by missing values [2], [3], [22], [32]. Bakir et al. [33] identified missingness in the data sets used in studying the effect of data homogeneity on software cost estimation in the embedded systems domain. 'Missing' is defined as not able to be found because a value is present but not in its expected place, or is not present or included when it is expected. The notion of incompleteness is broader, and can be explained as not complete or finished, imperfect. It may refer to lacking a part or parts; not whole; not full. Incompleteness is relevant to ESE in that small data sets might mean a model is not statistically significant, or lacks sufficient power.

Imputation is one of the procedures that have been used in dealing with the problem of missingness. Imputation is the "filling-in" of missing values with one or more plausible values. Khoshgoftaar & Van Hulse [34] conducted a comprehensive study of imputation techniques using real-world software measurement data sets. They considered the occurrence of missing values in multiple attributes, and compared three procedures: Bayesian multiple imputation, k-nearest neighbor imputation, and mean imputation. Their experimental results demonstrated that Bayesian multiple imputation is an effective imputation technique.

Multinomial logistic regression (MLR) was employed by Seliya & Khoshgoftaar [35] in imputing categorical missing values in the International Software Benchmarking Standards Group (ISBSG) multi-organization repository. Comparisons of MLR with other techniques for handling missing data, such as listwise deletion (LD), mean imputation (MI), expectation maximization (EM) and regression imputation (RI), under different patterns and percentages of missing data, showed the relatively high effectiveness of the MLR method [35]. Bayesian multiple imputation and regression imputation were found to be the most effective noise-resistant imputation techniques according to a comparative analysis reported by Van Hulse & Khoshgoftaar [36]. The presence of noise had a substantial impact on the effectiveness of the imputation techniques [34], [37].

Two embedded strategies (missing data toleration and missing data imputation) to handle missing data when using naïve Bayes and EM (Expectation Maximization) algorithms for software effort prediction were proposed by Zhang et al. [38]. The missing data toleration strategy simply ignores missing values and makes use of observed values of software projects for prediction. It has the advantage of low computational complexity. The missing data imputation strategy uses observed values of variables to estimate missing values. Experimental results drawn from their analyses of the ISBSG and CSBSG data sets demonstrated that both strategies outperformed classic imputation techniques.

Imputation techniques are a useful solution to the problem of missingness when the problem is not extensive, and researchers and practitioners must still adopt the most appropriate imputation techniques to resolve the varied conditions they might encounter, such as numerical or categorical values; class or attribute missing values; single or multiple missing attributes. Assessing the strength of imputation techniques under various conditions should enable researchers and practitioners to use the most suitable technique in each case.

*4) Inconsistency*

Inconsistency refers to a lack of harmony between different parts or elements; instances that are self-contradictory, or lacking in agreement when it is expected. In software engineering data sets it is essential that data values match the variables against which they have been recorded and can be clearly explained. Data must be appropriately recorded so as to ensure the integrity of any derived models.

Inconsistencies have been noted in prior studies of software measurement data [4], [32], [39]. Tan et al. [39] discovered inconsistencies when they studied a data set said to represent productivity trends in incremental and iterative software development. These inconsistencies included size and effort data that did not match from report to report. Therefore, they had to rework the reports and refer to the development team to resolve these mismatches, in order to arrive at an appropriate data set. Chen & Cheng [32] identified inconsistencies, in the form of unexplained questionable and null values, in a NASA data set employed in a discretization study intending to aid data accuracy. In their productivity analysis of a large data set [2], Liebchen and Shepperd found that size was recorded using different measures; in lines of code (LOC), in function points (FP), or in both, which made it difficult to compare projects in terms of this attribute. In their sensitivity analysis of data quality metadata [40], Fernández-Diego et al. discovered Lines of Code (LOC) measures that were not able to be explained. The affected projects were therefore removed. Inexperienced measurers were identified as contributors of poor data quality in the form of inconsistencies [41].

Bettenburg et al. [42] used a survey and machine learning to predict the quality of bug reports. Problems identified in bug reports included poor use of language (ambiguity) and bug duplicates, as well as incomplete information. In their study of software cost modeling Zhihao et al. [43] encountered suspicious repeated entries in one of the data sets which suggested data entry errors or too few examples for generalization. Projects that exhibited this anomaly were removed from the data set.

Consistency in meaning of data labels and recorded data values is one of the essential factors in achieving good quality data. As a community we use terms that we assume are collectively understood, but we do not have an ontology of issues that might better support shared understanding. ESE researchers and practitioners should ensure that the variables and values in data sets are easy to explain (and *are* explained), and should put in place mechanisms (e.g. variable definitions, range-checking for values) to resolve problems associated with the recording and interpretation of data.

*5) Redundancy*

Redundant and duplicate data in ESE data sets [8], [42] might lead to misleading results and can also detrimentally affect the performance of classifiers. Anh et al. [44] identified redundant data in the stakeholder name and nickname fields in their study of human factors for predicting issue lead time in open source projects. Prifti et al. [45] found that in their analysis of the Firefox bug repository there were 748 bugs that had been described in two or more groups when they applied a method that detected duplicates through local references. If effort modeling is based on such data then clearly there is scope for over-estimation of the actual effort required. Moreover, the building of classification models using data mining methods will be slowed by the additional processing needed to parse and consider the redundant values.

Another facet of redundancy that can lead to modeling problems is input data item dependence, or more generally, multicollinearity. Chidamber et al. [46] discovered that certain object-oriented metrics (Response for a Class, Weighted Methods Per Class and Coupling Between Object Classes) were highly correlated and so suggested that a subset be used in a linear regression model, otherwise the results generated could be unstable and difficult to interpret. The results obtained by El Eman et al. [47] when they studied the confounding effects of size on the validity of object-oriented design metrics confirmed that many of the metrics are highly correlated with class size.

Several ESE data sets, along with many automated data collection tools and environments, contain numerous measures. It is important that researchers and practitioners check for, and adjust for, any multicollinearity among these variables prior to their use in modeling.

### B. Relevance

The Oxford English Dictionary defines relevance, our second major class of quality attributes, as the quality or fact of being relevant – bearing upon, connected with, pertinent to, the matter in hand. In the context of ESE, relevance refers to having and using appropriate data to develop a model, normally for the purpose of classification or prediction. For instance, it would be inappropriate to use data collected from real-time systems development to build a model to predict development effort for banking systems. Relevance highlights the importance of the characteristics of the data being used in modeling, and an extensive body of literature has considered in particular the utility of single company data sets or multi-organization data sets in this regard.

*1) Heterogeneity*

Models generated from heterogeneous multi-organization data sets have been employed in estimating effort or predicting defects of software projects in a single company [42], [43], [48–51]. Results to date have been inconclusive as to whether single organization data sets are superior to those representing multiple organizations. Kocaguneli et al. [52] proposed the use of relevancy filtering when generating estimates using data from another project in their study of when to use data from other projects for effort estimation. Their results demonstrated that the use of cross-organization data usually results in estimation accuracies as high as those achieved through the use of within-organization data, provided that a relevancy filter is applied to the data prior to making estimates. Bakir et al. [33] studied the effect of data homogeneity on software cost estimation in the embedded systems domain and observed that all estimators performed better when they are trained on cross-domain data sets than when trained only on the within-domain (embedded software) data sets. The conclusion was that cross-domain data sets should be used for training estimators in embedded software cost estimation. (We note here that the single-company/multi-organization distinction may be in itself an over-simplification – some single organizations undertake hugely diverse projects.)

*2) Amount of Data*

The amount of data available for model building contributes to the likely statistical significance of generated models and so is another factor of relevance in terms of goal attainment. Small data sets are an acknowledged problem in ESE as they do not lend themselves to the generalization of results. The range of suitable analysis techniques is also constrained [53], [54] as some approaches assume the availability of a certain volume of data. Naturally, this issue is particularly pertinent to organizations that are just beginning a measurement programme, or that embark on projects that are substantially different to those undertaken in the past. To retain as much data as possible for software development effort modeling, Deng & MacDonell [55] employed an approach that systematically addressed the formalisation of data sets and employed domain-informed refinement to achieve a final usable data set drawn from the ISBSG repository.

Data pre-processing can also affect the amount of data available for modeling. A data set might initially comprise a large number of data points, but the application of stratification schemes or feature set selection strategies could result in data (sub)sets with too few data points to support significance testing. It is imperative for researchers to ensure that the pre-processing of data sets does not produce data subsets that lead to questionable generalizations of results, because the data sets are too small and/or because the modeling methods employed are not appropriate for the number of data instances.

The results obtained when Bakir et al. [33] assessed the effect of training data set size on the prediction performance of software cost estimation models was not consistent across all data sets used, which led to the conclusion that optimum training data set size depends on the method or algorithm used as well as the quality of data. Scarcity of data was noted as a great challenge to software engineering by Abrahamsson et al. [30] because it has serious implications for model validation and generalization. The leave-one-out (LOO) cross-validation procedure, often employed as a remedy to the limited data set size problem, was used when they [30] iteratively predicted development effort using an incremental approach. Naive Bayes and Random forest algorithms have also been proposed to increase the performance of prediction models based on small data sets and large data sets, respectively [56], [57].

*3) Timeliness*

The timeliness or currency of data is another issue of potential concern in regard to relevance, although it has received only limited attention to date in the ESE research literature. A survey of the data sets employed in software effort modeling reported by Mair et al. [58] in 2005 noted that many studies relied on data sets collected decades earlier. A perusal of ESE conference and journal publications today reveals ongoing use of these data sets. The characteristics of data sets should be constantly reviewed to ensure that changes in context and operation do not significantly reduce the relevance of the data to contemporary settings. While there is nothing inherently 'wrong' with the data sets in themselves, questions might well be asked about the appropriateness of the data in the context of present-day software development practice. If the intent is to build models for current use then data collected more recently would generally be preferable, other issues notwithstanding.

Timeliness has an additional impact in the context of real software development in that data are accumulated over time, as activities and projects are completed. In [59], MacDonell and Shepperd demonstrated empirically that failure to take the temporal nature of data accumulation into account leads to unreliable estimates of development effort. While much ESE research utilizes 'complete' data sets this represents an artificial scenario. Moreover, few such data sets comprise records of time. This is a characteristic that needs to be included in future data collection endeavors if the data are to be used in good faith – which leads usefully into our third class of quality issues.

## C. Provenance

The Oxford English Dictionary defines provenance as the fact of coming from some particular source or quarter; origin, derivation. It also refers to the history of ownership of a valued object or work of art or literature. Assurance of provenance, while especially significant for such objects, is also important in relation to digital artifacts or results that are generated by scientific applications. Information regarding provenance constitutes the audit trail, the proof of correctness of scientific results and in turn, can directly influence the extent of trust one might place on those results. For these reasons, the provenance of a scientific result is typically regarded to be as important as the result itself [60].

Considered broadly, provenance is related to the issue of experimental replication. Replications play a key role in empirical software engineering (as they do in other fields) by enabling the community to build cumulative knowledge about which results or observations hold under which conditions. Shull et al. [61] highlighted the importance of producing adequate documentation for an experiment to allow for replication. This echoes remarks made previously by Wieczorek [62], who indicated that few empirical software engineering studies were replicated, and even when the same data sets were used across different studies the results were not always comparable because of different experimental designs. She contended that studies in the ESE domain were difficult to replicate because they were not reported in a form that allowed for such comparisons. While this observation was made in 2002 the problem has persisted: in their replicated study of cross-company and single-company effort models using the ISBSG database Lokan and Mendes were unable to employ the same experimental procedure used previously because the procedure had not been fully documented [63]. Provenance systems have the potential to provide a more effective way of supporting replication as well as providing transparency regarding discrepancies in the results obtained from a replicated study and an original study.

### 1) Commercial Sensitivity

Commercial sensitivity is one of several constraints on provenance in ESE. Organizations that hold data that they believe gives them competitive advantage might not be willing to release the data to independent researchers, for fear of proprietary data being accessible to competitors. Similarly, they may be reluctant to release data if they believe it could be used to portray them in an unfavorable light. Even if researchers are able to have access to data, they are often required to sign non-disclosure agreements which prevent them from publishing the data with their results [2], [58], thus rendering such studies non-replicable.

The COCOMO-II data used in the pruning experiments of Zhihao et al. [43] was not published because it was collected on condition of confidentiality with the companies that supplied the data. Zhihao et al. [43] also did not disclose the details of locations, tasks and projects for the NASA data set due to confidentiality; however they provided general information about the data. While such precautions do lend some protection to the organizations involved, it has a consequent effect of limiting what can be learnt from the data analysis.

### 2) Accessibility

In the defect prediction study of Turhan et al. [49], they found it difficult to access failure logs because several large teams of contractors were working on projects for a single organization – NASA – and each viewed the failure logs as corporate critical. The authors note that acquisition of even coarse-grained information took years of careful negotiation. When finally provided, the data were highly sanitized by NASA meaning that the research team was not able to access project or module names. Robles [64] assessed the potential replicability of the experiments reported in papers published in the proceedings of the Mining Software Repositories Workshop/Conference between 2004 and 2009. It was discovered that only six out of 154 experimental papers were replicable because the data used in the other 148 original studies were not accessible.

To the best of our knowledge, the only work on provenance in the ESE domain is the Davies et al. [65] study that considered the provenance of software entities, specifically, the introduction of an anchored signature method to determine the provenance of source code contained within Java archives. This method was demonstrated successfully using a case study of a proprietary e-commerce application drawn from the Maven2 Java library repository. Determination of the provenance of the source code in this case was made possible by the accessibility of the Maven2 library. In keeping with the intent of this paper we would assert that the provenance of data *about* the source code – perhaps captured in the form of metrics – is as important as the provenance of the source code itself.

As noted above, Mair et al. [58] investigated the nature and type of data sets being used to develop and evaluate software project effort prediction systems. They noted at that time (2005) that only about 60% of all data sets were in the public domain. While significant growth in open source development over the last decade has increased the availability of empirical data, its utility for ESE is variable given the diversity of systems and development practices, and there are also questions over its general credibility as a basis for model-building. In addition, the open source model of development (or any particular model, for that matter) may not map well to other contexts. Contribution to generally accessible repositories such as those provided by the ISBSG (www.isbsg.org) and PROMISE (www.promisedata.org), with relevant provenance information, should be encouraged so that more data is made available to support ESE research and software engineering practice. If academia and industry are able to collaborate effectively this should increase the availability of data sets in the public domain, contributing to the replication of studies and to practice improvement.

### 3) Trustworthiness

The SE field is known for innovative work that proposes new tools, models, techniques and so on, but we are often far less effective in our evaluation of those proposals. Glass et al. [66] analyzed the software engineering literature prior to 2002 and concluded that SE was narrow in its research approach, with the "Formulate" approach being the dominant practice and few studies using evaluation as a core research activity. We have found similar outcomes in more recent reviews [67]. The extent to which research results hold beyond the often limited evaluations conducted and/or reported is therefore unknown. While this clearly applies to tools, techniques and methods, it is equally applicable to prediction and classification models. Catal & Diri [56] performed several experiments to assess researchers' claims that their fault prediction models provided the highest performance, and some of the models were revealed to perform far less well when evaluated on public data sets. This may reflect problems with the models themselves; or it may again signal the extent to which models are tied to the underlying data.

Limited access means that ESE researchers are generally 'at arm's length' from the data source, and consequently we are left with little option but to work principally with secondary data. We therefore rely heavily on the people and systems used to collect and verify that data. Greater adoption of provenance systems should provide data users with useful knowledge about the origins of the data and may influence the trust that can be placed on that data. It should also enable data providers and data users to track any changes that the data has undergone (for instance, whether the data has been masked, anonymized or transformed in pre-processing), information that is vital to ensuring that models are built with integrity.

## 4. DISCUSSION

Through this study we reviewed 57 papers that had addressed data quality in some way and identified 74 data quality issues considered in these papers. Figure 2 shows the distribution of these studies in terms of the three classes in our taxonomy. The total number of data quality issues is more than the number of papers because some papers addressed multiple issues. It is quite evident that issues of accuracy have received the most attention (at 65%), followed by relevance (with 23%) and provenance (with just 12%).

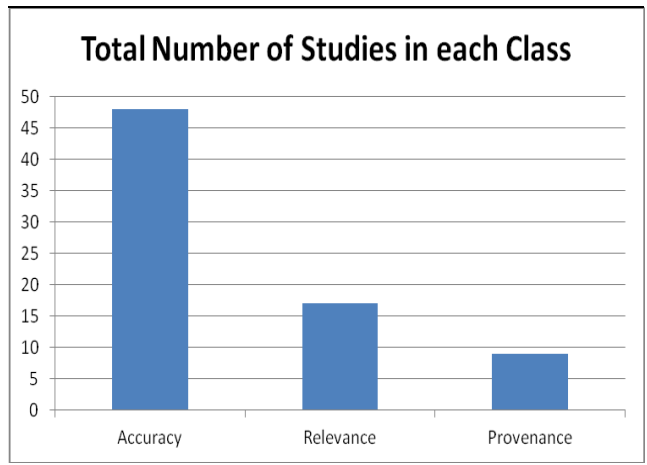

Figure 2. Distribution of Data Quality issues considered in ESE studies.

A breakdown of studies that considered accuracy issues is presented in Figure 3. Noise and incompleteness have received the most attention with each being considered in 27% of the studies reviewed in the accuracy class. Outliers followed in third position with 21%, then inconsistency with 17%, and finally redundancy which featured in 8% of the studies in this group. Note that even though data inconsistency was identified as an issue in some studies, mostly in relation to data preprocessing, it was not a main theme in any of the studies, unlike noise, incompleteness and outliers.

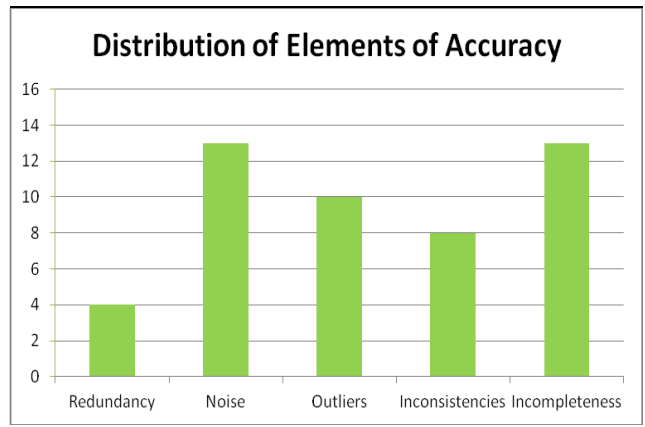

Figure 3. Distribution of elements of accuracy.

Relevance was the second most important class of the taxonomy in terms of issue coverage. We present a breakdown of the elements of relevance in Figure 4. Heterogeneity has received the most attention at 47% of studies reviewed in this class, a not unexpected result as there is still contention about whether to build models with cross-company data sets or single-company data sets. The size of the data set used in modeling followed with 41% coverage. Size of data sets is seen as a particular challenge in empirical software engineering because most of the data sets are small and so do not lend themselves readily to modeling, a situation further exacerbated when some of the data has to be discarded due to other data quality problems such as missingness. Few studies considered the timeliness of the data used (12%). We find this a little discouraging, considering the fact that many of the data sets, especially those in the public domain, are relatively old. The dynamic nature of software development practice demands new data to reflect current work practices.

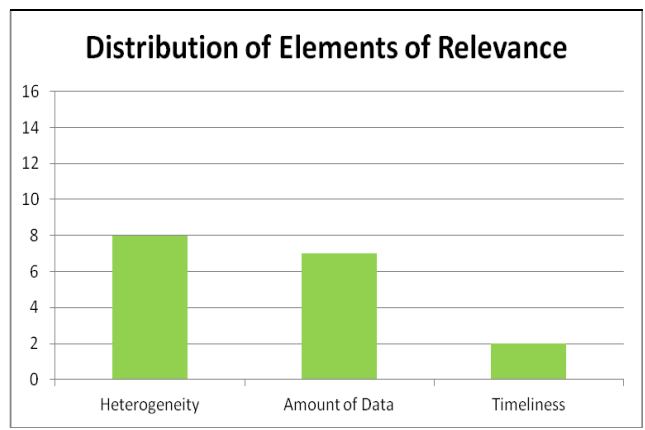

Figure 4. Distribution of elements of relevance.

The class of provenance was the least frequently addressed of those considered in the taxonomy. The distribution of the constituent elements of provenance is shown in Figure 5. Commercial sensitivity and accessibility were highlighted by approximately 45% each of the studies reviewed in the provenance class, and only 10% addressed the issue of trustworthiness.

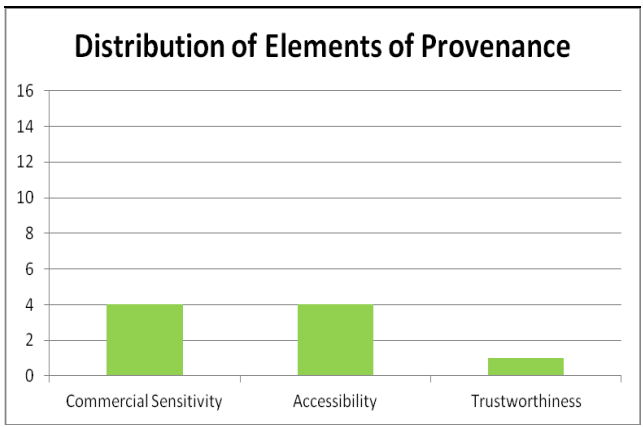

Figure 5. Distribution of elements of provenance.

Figures 3 to 5 depict the contributions of the constituent elements to their respective class of the taxonomy. In Figure 6, we show the contribution of each individual element to the taxonomy as a whole. Noise and incompleteness (missingness) each contributed approximately 18% to the total issue coverage as classified by the taxonomy. Outliers made up of 14%, inconsistency 11% and redundancy 5%. Heterogeneity contributed 11%, amount of data 9% and timeliness 3%. Commercial sensitivity and accessibility contributed 5% each and trustworthiness only made a 1% contribution. These results and their representation in Figure 6 illustrate the diversity and uneven consideration – let alone treatment – of the data quality challenges relevant to ESE research and practice.

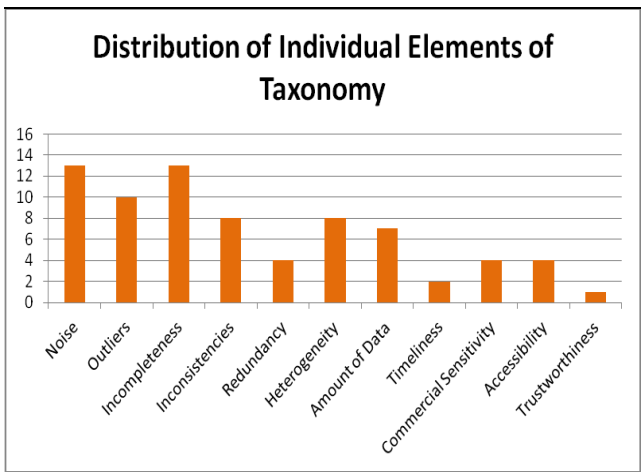

Figure 6. Distribution of individual elements of the taxonomy.

According to our review it appears that the issues of provenance and provenance systems have received very limited, and indirect, attention in empirical software engineering research (that is, papers have considered issues such as trustworthiness but not as an aspect of the broader topic of provenance). Devoting greater attention to provenance and systems that can support it has the potential to improve the entire procedure of data collection, generation and use in prediction and classification, facilitating the replication of studies. Given that science proceeds on the basis of evidence-based outcomes whose validity is assured through methods such as replication, it is imperative that researchers and practitioners devise means to make more data more available, to improve the broader practice of software engineering. Provenance has the potential to improve the reputation of ESE data and its use as, over time, we acquire a growing understanding of the processes and organizations that generate trustworthy data. To date this issue has attracted limited interest – the most obvious example being the quality category ratings assigned to observations recorded in the ISBSG repository. While this is certainly a very useful starting point, it should be noted that these ratings are assigned by ISBSG quality reviewers based on their assessment of the integrity of the data, rather than by anyone associated directly with the data's collection. Effective records of data provenance should also support the identification of likely problems with data (and any resultant models) because there will be provenance information against which abnormal results can be tracked.

Through our review it became clear that one of the reasons contributing to the weaknesses identified in ESE data sets is the inadequate reporting of data collection procedures – there seems to be no standard expectation that the data collection process be described (let alone the data be included), and so it is a minority practice in ESE research papers [68]. Even in the few papers in which collection procedures are reported, problems associated with data collection are generally not, making it difficult for ESE researchers to have a clear understanding of the causes of any data quality issues. The necessary but unquestioning use of secondary data has thus contributed to the inadequacy of efforts towards the resolution of data quality problems. The most appropriate solution to these problems is to prevent them from occurring at the point of data capture.

## 5. CONCLUSIONS

In this study we have reviewed the relevant literature to identify the major data quality issues in ESE in order to improve the community's awareness and understanding of the quality challenges (and current solutions) in ESE research and practice. Issues grouped under the class of accuracy have received the most attention from the research community, whilst provenance issues have received the least.

The potential of provenance to assure data quality has not been exploited in ESE. Adopted sensibly and systematically, provenance should increase the reputation and trustworthiness of the data that is used in modeling, which will consequently result in higher quality models. Although data quality has been considered from several perspectives in ESE, as yet we do not have sufficient evidence on how it influences the practice of software engineering as almost all the studies considered here are drawn from academic institutions – more extensive collaboration with industry is essential in order to understand and then improve data quality management in practice. No single study evaluated all the aspects of the taxonomy, indicating that the treatment of data quality is not a holistic endeavor in ESE. Genuine practice improvement will only be possible if this currently piecemeal situation is collectively addressed. To that end we are in the process of developing a provenance software tool to gather provenance data during the data collection and processing stages of software development projects. This tool will be able to capture and explain the causes of data quality problems at source and should then inform the development of suitable preventive measures.